# Magnetoelastic coupling and spin excitations in the spin-gap system (VO)$_2$P$_2$O$_7$: A Raman scattering study

M. Grove, P. Lemmens, and G. Güntherodt

*2. Physikalisches Institut, RWTH-Aachen, D-52056 Aachen, Germany*

B.C. Sales

*Oak Ridge National Laboratory, Oak Ridge, Tennessee 37831-6393, USA*

F. Büllesfeld and W. Assmus

*Physikalisches Institut, Universität Frankfurt, D-60054 Frankfurt, Germany*

(November 2, 1999)

Single crystals of the spin-gap system (VO)$_2$P$_2$O$_7$ have been investigated by means of Raman scattering. At temperatures below $T = 75$ K $\approx 2\Delta_{01}(\mathbf{k} = 0)$, with $\Delta_{01}$ the singlet-triplet gap, a shoulder appears at 47 cm$^{-1}$. An analysis of the polarization selection rules and the temperature dependence of the scattering intensity leads to its identification as a singlet bound state with very small binding energy. Therefore in contrast to recent theoretical models the importance of frustration for this compound must be rejected. The observation of a strong anharmonicity of three phonons with energies close to $\Delta_{01}$ at the zone boundary and additional quasielastic light scattering, however, both point to a strong spin-phonon coupling in this system. We propose this coupling to be the origin of the second bound triplet state observed in recent neutron scattering experiments.

75.10.Jm, 75.50.Ee, 78.30.-j

## I. INTRODUCTION

Low-dimensional quantum spin systems have attracted an intense theoretical and experimental attention over the last years. This interest is motivated by the occurrence of exotic ground states and unconventional excitation spectra in one- and quasi-one-dimensional quantum antiferromagnets. In low-dimensional systems quantum fluctuations and suppression of long-range order at finite temperatures play an important role. The low energy excitation spectrum of a one-dimensional antiferromagnetic spin-1/2 Heisenberg chain is therefore not given by spin waves but by a continuum of spinons[1,2]. These spinons carry spin-1/2 and therefore possess fermionic character[2,3]. The excitation spectrum is gapless.

There are different mechanisms that form a spin-gap in the excitation spectrum of such a system. One is to couple an even number of spin chains via an additional interchain interaction. This leads to a spin ladder which has attracted a lot of interest over the last years[4]. The dimerized spin chain with exchange coupling of alternating strength realizes a similar dimer-like spin configuration and the opening of a singlet-triplet gap ($\Delta_{01}$)[5]. The triplet branch that forms this gap consists of confined spinons. In the case of an additional attractive interaction magnetic bound states exist for energies $E \leq 2\Delta_{01}$. These singlet or triplet states are bound from triplets due to a generalized interchain interaction or a frustration of the spin system. In alternating spin chains or ladders frustration is a key parameter that determines the spectral weight of these states[6–9].

An important aspect of triplet-triplet interaction in "real" systems, however, is the often ignored spin-phonon coupling. Recent theoretical models show that it can establish bound states or soliton states of considerable spectral weight[10–16]. Dramatic effects occur especially if phonons exist with energies close to the magnetic excitation spectrum. In this case the magnetic properties are strongly renormalized compared to the static limit due to the coupled dynamics of spin and lattice degrees of freedom. In this paper we will show undoubtedly experimental evidence for this scenario in (VO)$_2$P$_2$O$_7$ and discuss its implication for the recently observed triplet bound states in neutron scattering experiments[17].

Initially, (VO)$_2$P$_2$O$_7$ (vanadyl pyrophosphate) was considered to be an excellent realization of a two-leg Heisenberg spin ladder[4,18,19]. Indeed, its crystal structure shown in Fig. 1 consists of ladders with double V-O-V legs along the $a$ axis direction. Along the $b$ axis direction chains with strongly alternating exchange coupling are formed. They are defined by an exchange coupling path via double V-O-P-O-V links through the phosphate groups alternating with V-O-V links between edge sharing VO$_5$ square pyramids[20]. The magnetic properties of (VO)$_2$P$_2$O$_7$ arise from the $s = 1/2$ V$^{4+}$ ions situated within distorted VO$_6$ octahedra.

In recent inelastic neutron scattering experiments the strongest (antiferromagnetic) dispersion with $\Delta_{01}(\mathbf{k} = 0) = 36$ K has been found for $\mathbf{k}$ along the $b$ direction[17]. The zone-boundary energy of this branch is $E_{ZB} = 180$ K. This strong exchange coupling parallel to the alternating chain direction excludes an interpretation of this compound as a spin ladder. In addition to this branch a second non-degenerate triplet branch is ob-served with $\Delta'_{01}(\mathbf{k}=0) = 69$ K. The latter energy is slightly below $2\Delta_{01}(\mathbf{k}=0)$. The second branch has consequently been interpreted as the triplet bound state of a strongly alternating chain system. This is the first experimental report of such a bound state in neutron scattering investigations.

Unfortunately, recent calculations have shown that the spectral weight of a triplet bound state in an alternating chain system is rather limited in $\mathbf{k}$ space to a region close to the border of the Brillouin zone. This at least leaves the need for an additional interaction[21]. In two publications the possibility of an additional diagonal frustration via the phosphate groups has been discussed to yield or enhance the spectral weight of the triplet bound state[22,23].

An alternative model motivated by NMR experiments is based on the very large unit cell (eight formula units) of $(VO)_2P_2O_7$ with four chains and two crystallographic non-equivalent V-O-V bonds. The two branches may therefore result from two different alternations of the chain systems A and B[24]. The two observed singlet-triplet gaps were attributed to $\Delta^A_{01} = 68$ K and $\Delta^B_{01} = 35$ K in good agreement with neutron scattering. These gaps correspond to alternations of $J_2/J_1 = 0.83$ and $0.67$ of the chain systems A and B, respectively. The corresponding difference of the V-O-V total bond length between the two chain systems, however, is only $0.02$Å (of a total bond length $d_{VOV} = 3.23$ Å). It is not clear up to now whether such a small crystallographic difference is sufficient to significantly distinguish the two proposed spin subsystems.

The remainder of this paper is organized as follows: In section II we describe the setup used for and in section III the results obtained from the Raman scattering experiments. The subsequent sections discuss these results and present the main conclusions.

## II. EXPERIMENTAL SETUP

We have performed Raman scattering experiments on well-characterized single crystals from two different batches[25,26] in quasi-backscattering geometry. The two batches of samples gave essentially the same experimental data. For the experiments we used the $\lambda = 514.5$ nm excitation line of an Ar$^+$ ion laser and a laser power below $0.2$ W/cm$^2$. The incident radiation did not increase the temperature of the samples by more than 1 K. The samples themselves were glued onto a copper sample holder and immersed in flowing He-exchange gas investigating a temperature range from 300 K down to 5 K. Scattered light was analyzed with an XY-Dilor Raman spectrometer and a back-illuminated CCD detector. For the experiments in an applied magnetic field a multi-pass Sandercock-type Fabry-Perot spectrometer with an avalanche diode detector was used.

## III. EXPERIMENTAL RESULTS

In the following several experimental facts demonstrating strong spin-phonon coupling and magnetic light scattering will be presented: first, a strong hardening of three phonons, second, a renormalization of the scattering intensity in the low frequency range, and finally, quasielastic scattering due to fluctuations of the energy density of the spin system.

### A. Anharmonic phonons

Numerous phonons have been observed in the frequency range up to 1200 cm$^{-1}$ in each polarization. This is in good agreement with a factor group analysis predicting 155 and 154 phononic excitations in A and B symmetry, respectively. The A symmetry components are resolved in $(aa)$, $(bb)$, and $(ab)$ light scattering polarization while B symmetry components are given in $(ac)$ and $(bc)$ light scattering polarization with $(\vec{E}_i, \vec{E}_s)$ the polarization of the incident and scattered light parallel to $a$, $b$, or $c$, the crystal axes of the lattice. Investigating the frequency and halfwidth of these modes we notice an anomalous strong anharmonicity of certain phonons as a strong hardening in energy and a decrease in line width with decreasing temperature. Only phonons with an energy lower or comparable with the maximum energy of the triplet dispersion branch $(E_{ZB} = 180 \text{ K} \approx 125 \text{ cm}^{-1})^{17}$ show these pronounced effects. This energy scale is definitely magnetic. Therefore these anomalies are attributed to spin-phonon coupling.

In Fig. 2 Raman spectra of the A symmetry phonon at 123 cm$^{-1}$ are presented in dependence on temperature. Similar results were obtained for an A symmetry phonon at 70 cm$^{-1}$ and a B symmetry phonon at 118 cm$^{-1}$ see Fig. 3. These modes show a strong hardening with decreasing temperature of 10% and 5%, respectively. The line shapes of these phonons have a symmetric Lorentzian form and show a drastic decrease in halfwidth by a factor of two with decreasing temperature.

While the integrated intensity of the phonons in $(bb)$, $(cc)$ and $(ac)$ polarization is independent of temperature the interchain $(aa)$ polarization appears to be peculiar. In Fig. 4 it is shown that the mode at 123 cm$^{-1}$ is strongly damped in interchain $(aa)$ polarization for temperatures above 100 K. Its intensity decreases exponentially towards higher temperatures. The assumption that a thermal population of the triplet branch is responsible for this drop of intensity leads to $I \propto 1 - \exp(-\Delta/T)$ with a gap of $\Delta = \Delta_{01}(\mathbf{k}=0) = 36$ K determined by a fit (straight line in Fig. 4).

To recapitulate, all observations concerning anharmonic phonon effects correlate with the energy scale of the spin system as given by the dispersing triplet branch $\Delta_{01}(\mathbf{k})$ or $2\Delta_{01}(\mathbf{k}=0) = 72$ K $\equiv 50$ cm$^{-1}$. The latter energy is relevant if the phonons couple to a magnetic two-particle excitation. More precisely, the hardening of the phonon frequencies upon cooling saturates below a temperature $T = 75$ K. Furthermore, only phonons with an energy comparable to or smaller than $E_{ZB}$ are involved. Therefore, we conclude that these anomalous anharmonic effects are due to a strong spin-phonon coupling. For an overview we refer to Tab. I giving the anharmonic effects for different polarizations.

### B. Magnetic bound states

Magnetic exchange light scattering is observed as a shoulder in the scattering intensity at $47 \pm 3$ cm$^{-1}$, an energy just below $2\Delta_{01} = 50$ cm$^{-1}$. This intensity is only observed in $(aa)$ and $(bb)$ polarization and develops for low temperatures as shown in Fig. 5 (a) and (b). It is more than two orders of magnitude smaller than the phonon light scattering intensity. Furthermore, it is not observed in any other polarization, e.g., in the crossed $(ab)$ polarization shown in Fig. 5 (c).

There is a pronounced linear increase of this scattering intensity for $T < 70$ K $\approx 2\Delta_{01} = 72$ K. In light scattering polarization parallel to the chain direction $(bb)$ this increase is divided in two temperature regimes: first, from 72 K down to 25 K the intensity slowly increases and second, below 25 K the intensity strongly increases with a linear dependence on temperature which is shown in Fig. 6. In $(aa)$ polarization the intensity of the shoulder increases with a linear dependence on temperature for $T < 50$ K. We neither observe a magnetic field dependence in $(aa)$ nor in $(bb)$ polarization for magnetic fields up to 6 T as presented in Fig. 7 for $T = 2.4$ K.

Summarizing, we observe an additional shoulder in $(aa)$ and $(bb)$ polarization at an energy just below $2\Delta_{01}$. In crossed $(ab)$ configuration we observe no contribution of magnetic light scattering (see Fig. 5 (c)). This would be the preferred polarization for conventional 2D (square plane) antiferromagnets[27]. In the following we will refer to this intensity of the shoulder as singlet bound state scattering as discussed further below.

### C. Quasielastic scattering

Additional quasielastic scattering intensity in the very low frequency range can only be observed in $(aa)$ polarization at high temperatures. To demonstrate how the quasielastic scattering changes with temperature, we show in Fig. 8 typical spectra obtained at 40, 55, 75, and 200 K. The quasielastic scattering is clearly observed in a temperature range from 300 K down to 50 K. Rayleigh scattering appears only below $|\omega| < 20$ cm$^{-1}$ and therefore is narrow enough not to mask the quasielastic scattering in the present experiments. The scattering intensity observed for $T \leq 50$ K is very weak, as can be seen in the spectrum obtained at 40 K.

## IV. DISCUSSION

Quasielastic scattering can be described in two different models: first it could be assigned to spin diffusion processes or second to fluctuations of the energy density of the spin system. The latter process is especially important if the magnetic system is coupled to the lattice[28]. This mechanism has been identified and successfully used to describe quasielastic scattering, e.g., in spin-Peierls systems[34]. Here spin-phonon coupling leads to a spontaneous structural transition connected to a dimerization of the spin system.

In the first case the low frequency Raman spectra would be Gaussian-like[29,30]. Therefore, we have fitted the observed spectra with the following Gaussian-type spectral function, using the method of least squares:

$$I(\omega) = \frac{K_G^2}{\sqrt{2\pi}\Gamma_G} \exp\left(-\frac{\omega^2}{2\Gamma_G^2}\right) + \text{background}, \quad (1)$$

where $\Gamma_G$ and $K_G$ in Eq. 1 are the damping constant and the coupling coefficient, respectively. In the latter case the low frequency Raman spectra would have the form of a Lorentzian curve[31,32]. For fitting the spectra we use the following Lorentzian spectral function applying the same method as above:

$$I(\omega) = \frac{K_L^2 \Gamma_L}{\omega^2 + \Gamma_L^2} + \text{background}, \quad (2)$$

where $\Gamma_L$ and $K_L$ correspond to $\Gamma_G$ and $K_G$ in Eq. 1.

We have measured the temperature dependence of the low frequency Raman spectra in $(aa)$ polarization. Particularly, as we show in Fig. 8, the observed spectra fit better to the Lorentzian spectral function in Eq. (2) than to the Gaussian spectral function in Eq. (1). In the inset of Fig. 8 we show as an example Raman spectra obtained at 200 K in comparison with proper Gaussian and Lorentzian curves. Especially for energies larger than 45 cm$^{-1}$ the deviation between the Gaussian curve and the experimental data is clearly visible.

The energy of a magnetic system is not constant but fluctuates about the average determined by the lattice temperature. Spin-phonon coupling decreases the time scale and therefore increases the spectral weight of the fluctuations[28]. In the previous chapter III A we have demonstrated strong anharmonic behavior of several phonons whose energies are comparable with the energy scale of the spin system. This strongly suggests that the quasielastic scattering is due to fluctuations of the energy density of the spin system.

In this case one can easily deduce the magnetic specific heat $C_m$ from the quasielastic scattering intensity. This has already been done for KCuF$_3$[33] and CuGeO$_3$[34].

According to the theory of Reiter[28] and Halley[32], the intensity of the quasielastic scattering is given by the following Fourier component of the correlation function of the magnetic energy density:

$$I(\omega) = \gamma \int_{-\infty}^{\infty} dt \exp^{-i\omega t} \langle E(\mathbf{k},t)E^*(\mathbf{k},t) \rangle, \qquad (3)$$

where $E(\mathbf{k},t)$ and $\hbar\omega$ denote the magnetic energy density with the scattering vector $\mathbf{k}$ and the energy transfer, respectively. $\gamma$ is a $\mathbf{k}$- and $\omega$-independent but polarization dependent coefficient. We assume that $\gamma$ is temperature independent. Introducing the hydrodynamic form given by Halperin and Hohenberg[31] for the correlation function $\langle E(\mathbf{k},t)E^*(\mathbf{k},t) \rangle$ and using the fluctuation-dissipation theorem, one obtains the expression

$$I(\omega) = \frac{\gamma\omega}{1-e^{-\hbar\omega/k_BT}} \frac{C_m T D_T k^2}{\omega^2 + (D_T k^2)^2}. \qquad (4)$$

The thermal diffusion coefficient $D_T$ is given by

$$D_T = K/C_m, \qquad (5)$$

where $K$ is the magnetic contribution to the thermal conductivity. When $\omega \sim D_T k^2 \ll \omega_0$, one obtains $\mathbf{k} \simeq \mathbf{k_0}\sin\theta/2$ for transparent crystals, where $\theta$ is the scattering angle of the light and $\omega_0$ and $\mathbf{k_0}$ characterize the incident light. For non-transparent crystals the momentum transfer is determined by the penetration depth but will be of the same order of magnitude as for transparent ones. In a high temperature approximation with $\hbar\omega/k_BT \ll 1$ Eq. 4 is simplified to:

$$I(\omega) = \frac{k_B\gamma}{\hbar} \frac{C_m T^2 D_T k^2}{\omega^2 + (D_T k^2)^2} \qquad (6)$$

Using Eq. 2 and Eq. 6 we can estimate the magnetic specific heat from the integrated scattering intensity. The assumption of a constant background will not be altered by Rayleigh scattering, because as already shown in section III C even at low temperatures elastic scattering is only observable below $|\omega| < 20$ cm$^{-1}$.

The magnetic specific heat derived from the quasielastic scattering is shown in Fig. 9 as a function of temperature. The observed maximum at $T_{max} = 55$ K is in good agreement with theory. DMRG calculations[35] for an alternating chain with a coupling constant $J = 110$ K and a dimerization $\delta = 0.09$ give a maximum at $T = 52$ K (parameters from neutron scattering experiments in Ref. 17). These results are shown in Fig. 9 as solid curve. It should be noted, however, that the maximum in the specific heat mainly depends on the exchange coupling parameter and only weakly on the dimerization of the spin chain. Therefore, a similar reasonable agreement is found using the model of Bonner and Fisher[36] for a homogeneous spin chain with nearest-neighbor (nn) intrachain exchange interaction $J_{eff}$. The expected maximum in the magnetic specific heat at $T_{max} = 0.64 \cdot J_{eff}$ leads to $J_{eff} = 90$ K. This exchange coupling constant is still in qualitative agreement with neutron scattering experiments for $\mathbf{k}$ along the chain direction[17]. As the sensitivity of the magnetic specific heat to the dimerization is only weak the results presented here can only partially be used to distinguish between models that use a single or two different alternations. At least the importance of spin-phonon coupling can be highlighted together with a fixing of the energy scale of the spin system. In this sense the neutron scattering results with a dominant exchange path in $b$ axis direction are confirmed. The magnitude of spin frustration, if existing in $(VO)_2P_2O_7$, should not be too large as it should shift the maximum in the specific heat strongly and should prevent a reasonable agreement between the Bonner-Fischer model and DMRG calculations. This has been shown for the frustrated spin-Peierls system $CuGeO_3$[34].

In the previous chapter magnetic light scattering with a shoulder at $47\pm3$ cm$^{-1}$ in $(aa)$ and $(bb)$ polarization has been described. The identification of this signal as a singlet bound state is based on its negligible magnetic field dependence (see Fig. 7) and the linear rise in scattering intensity for decreasing temperature (see Fig. 6). This behavior can be well compared to the singlet bound state observed in $CuGeO_3$[37]. The small binding energy of this signal just below $2\Delta_{01} = 50$ cm$^{-1}$ would only be expected for a singlet bound state in a spin chain with negligible frustration[8]. The weak spin-orbit coupling evidenced in ESR experiments[26] in addition excludes one-magnon scattering, e.g., from the upper dispersing branch $(\Delta'_{01})$[17] in $a$ direction. We neither observed magnetic scattering corresponding to the gap $2\Delta^A_{01} = 95$ cm$^{-1}$ postulated from NMR experiments[24] nor a two-magnon scattering signal from the zone-boundary energy of the lower triplet branch $\Delta_{01}$. This latter signal may only be expected for a two-dimensional antiferromagnet.

The reason for the observation of two temperature regimes in the scattering intensities of the singlet bound state in $(bb)$ polarization has to be further discussed. The ground state in $(VO)_2P_2O_7$ is a singlet ground state with a pre-shaped spin-gap due to the static dimerization of the alternating chain. Experimentally the intensity of the bound state exists only for temperatures $T < 2\Delta_{01} = 72$ K and it is continuously increasing with lowering temperature, as the number of thermally induced triplet states is decreasing (we refer to Fig. 6). A further increase of intensity is observed in $(bb)$ polarization pointing to a second energy scale involved. This scale is most probably connected with magnetic or magnetoelastic interchain interaction. If spin dimers on adjacent chains are correlated a further increase of the spectral weight is induced. A quasi-two dimensional dimer pattern is formed for temperatures below the "coherence" temperature $T_{coh} = 25$ K given by the change of slope in Fig. 6. To compare this temperature with an energy scale of the system the two-particle nature of the scattering processes has to be taken into account giving a coupling energy $E_{coh} = 12.5$ K. Estimates of the interchain coupling constant based on neutron scattering results range from a weak ferromagnetic to an appreciable antiferromagnetic coupling, i.e. $J_\perp = -2.4 - 25$ K[17,22]. The scattering intensity in the $(aa)$ polarization as shown in Fig. 6 is not influenced by coherence as interchain exchange processes are involved. Consequently, the onset of this intensity contribution is observed at lower temperature compared to the intrachain ($bb$) polarization.

It is evident that the situation we face in $(VO)_2P_2O_7$ is completely different from a spin-Peierls compound, e.g., $CuGeO_3$, where the spin-gap $\Delta_{01} = 24$ K exists only for $T < T_{SP} = 14$ K and the interchain coupling $J_\perp \approx 15$ K is larger than $T_{SP}$. Therefore, in $CuGeO_3$ similar crossover phenomena are only observable in substituted samples[37].

The central question in understanding $(VO)_2P_2O_7$ is the interpretation of the second triplet mode observed in neutron scattering or NMR experiments. The identification of this compound as a strongly alternating spin chain system as a starting point is unquestionable. However, this model does not allow triplet bound states of considerable spectral weight throughout the Brillouin zone. As stated above, two possible explanations of the second triplet mode have been discussed so far, i.e. a magnetic bound state due to an additional frustration or the existence of two differently alternating spin subsystems that should give rise to two separate triplet modes. We are going to give a short review about these ideas and compare these results with our experimental observations.

A frustration mediated through a diagonal AF exchange $J_\times$ across the phosphate plaquette corresponding to the oxygen mediated superexchange pathways in the $ab$ plane was proposed by Uhrig et al.[22] and by Weiße et al.[23]. This 2D frustrated model qualitatively describes the full dispersion of the triplet bound state with $\mathbf{k}$ both in $b$ direction as well as in $a$ direction[17]. Spin frustration as an attractive interaction in alternating chains should give rise even more to a large binding energy of singlet bound states. These states were found in the calculations at $\Delta_{00} = 1.66 \cdot \Delta_{01} \equiv 41.4$ cm$^{-1}$ and $\Delta'_{00} = 1.94 \cdot \Delta_{01} \equiv 48.6$ cm$^{-1}$.[23] In addition, the spectral weight of the first singlet bound state should be larger than that of the second one. However, as shown above there is no experimental evidence in Raman scattering experiments for magnetic Raman scattering at frequencies smaller than 45 cm$^{-1}$. Therefore, in the following we use the hypothesis that the large spectral weight of the second triplet branch is not due to frustration or interchain interaction.

The recently proposed model of two spin subsystems based on NMR experiments[24] and (at least) two crystallographic inequivalent V chains would lead to the identification of the observed singlet bound state at 47 cm$^{-1}$ as originating from $\Delta^B_{01} = 35$ K. The second independent spin subsystem with $\Delta^A_{01} = 68$ K should lead to a similar signal close to or below 95 cm$^{-1}$. In our Raman experiments we have found no evidence for additional magnetic scattering in this frequency range. Therefore this model is not supported by our results.

In addition, a more general problem not addressed in Ref. 24 is the nearly doubling of the gap energy going from the first to the second spin subsystem. As these are believed to originate from a small crystallographic inequivalence this factor of approximately two is questionable to be accidental. A weak binding, sufficient to decrease the energy of the bound state below the continuum and thereby enhancing its spectral weight, would, on the other hand, be consistent with the observed factor of two.

A probably missing ingredient to solve this puzzle is strong spin-phonon coupling[10–16]. Theoretical approaches cited here are based either on discussing bound state phenomena or on non-adiabatic effects. Indeed, we observe a strong anharmonic behavior of three phonons, whose energies are the same as that of the spin excitations. The condition $\omega < \Delta_{01} = 25$ cm$^{-1}$ for an adiabatic approach is not fulfilled for these important modes, either. The magnetic properties of $(VO)_2P_2O_7$ are therefore strongly renormalized due to the coupled dynamics of spin and lattice degrees of freedom compared to the static limit of an alternating spin chain. In Ref. 10 it was highlighted that in such a case spin-phonon interaction becomes important for the existence of well defined singlet-triplet excitations. Furthermore, it was shown that phonons with an energy smaller than or close to the maximum energy of the triplet dispersion branch $\Delta_{01}$ show anharmonic behavior due to damping by two-magnon processes[38]. As already shown above the damping of the phonon mode in $(aa)$ polarization is only connected with $\Delta_{01}$. No effect of the second proposed spin system could be found.

Infrared absorption spectra[39] of quite a number of pyrophosphates that share the existence of a phosphate plaquette have characteristic phonons in the frequency range of 100-130 cm$^{-1}$, for example in $Ba_2P_2O_7$ at 118 cm$^{-1}$. The phonon modes at 70 cm$^{-1}$, 118 cm$^{-1}$, and 123 cm$^{-1}$ in $(VO)_2P_2O_7$ can therefore be assigned to the phosphate plaquette. From the compound $VODPO_4 \cdot \frac{1}{2}D_2O$, a precursor of $(VO)_2P_2O_7$, it is known that the spin dimer is formed over the phosphate plaquette[40]. Strong spin-phonon coupling of these modes is obviously modulating the exchange over the phosphate plaquette.

Therefore, the stated non-adiabaticity of the spin and the lattice system may explain the second triplet bound state observed in neutron scattering and the singlet bound state observed in our experiment.

## V. CONCLUSION

Using Raman light scattering we have shown strong experimental evidence for a non-negligible spin-phonon coupling in the compound $(VO)_2P_2O_7$. This evidence is based on the observation of a strong anharmonicity of certain phonons whose energies are equal to the dispersing triplet branch, a negligible binding energy of a singlet bound state and strong quasielastic scattering due to fluctuations of the energy density of the spin system. From the quasielastic scattering we have de-duced the magnetic specific heat $C_m$ with its maximum at $T_{max} = 55$ K. As an appropriate model for the spin-gap system $(VO)_2P_2O_7$ we propose an alternating spin chain with strong spin-phonon coupling.

## VI. ACKNOWLEDGMENTS

We thank G. Uhrig, A. Klümper and R. Raupach for stimulating discussions and G. Els for experimental help. This work was supported by DFG through SFB 341 as well as by INTAS 96-410.

TABLE I. Polarization dependence of the anharmonic behavior (x) of the phonons at 70 cm$^{-1}$, 118 cm$^{-1}$, and 123 cm$^{-1}$. A dash (-) denotes no observation of anharmonic behavior in this light scattering polarization.

| phonon frequency | polarization | | | | | |
|---|---|---|---|---|---|---|
| | (aa) | (bb) | (cc) | (ab) | (ac) | (bc) |
| 70 cm$^{-1}$ | - | - | x | - | | |
| 118 cm$^{-1}$ | | | | | x | x |
| 123 cm$^{-1}$ | x | x | x | x | | |

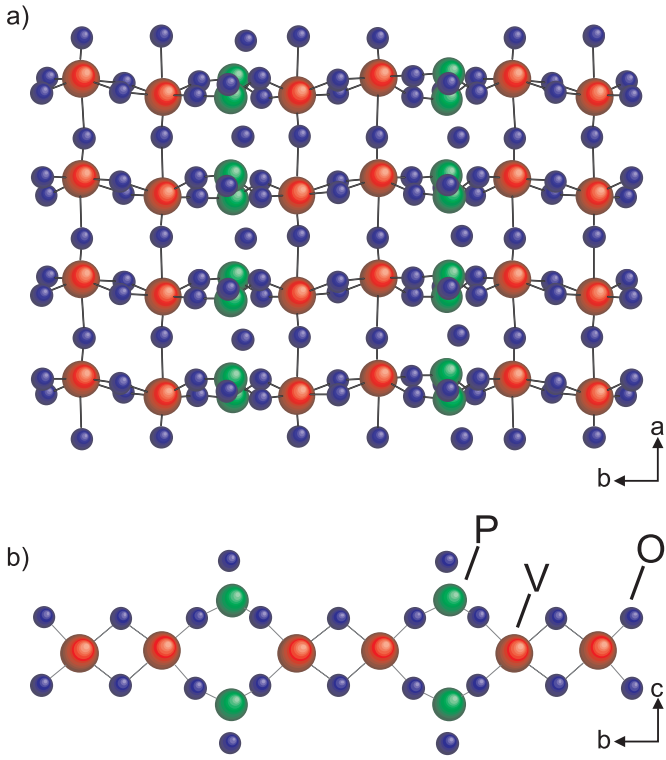

FIG. 1. The schematic projection of the crystal structure of $(VO)_2P_2O_7$ in the ab and bc plane showing a) the "ladder" along the $a$ axis direction and b) the alternating chain along the $b$ axis direction, respectively. V-ions are shown in gray, P in light gray and O in black.

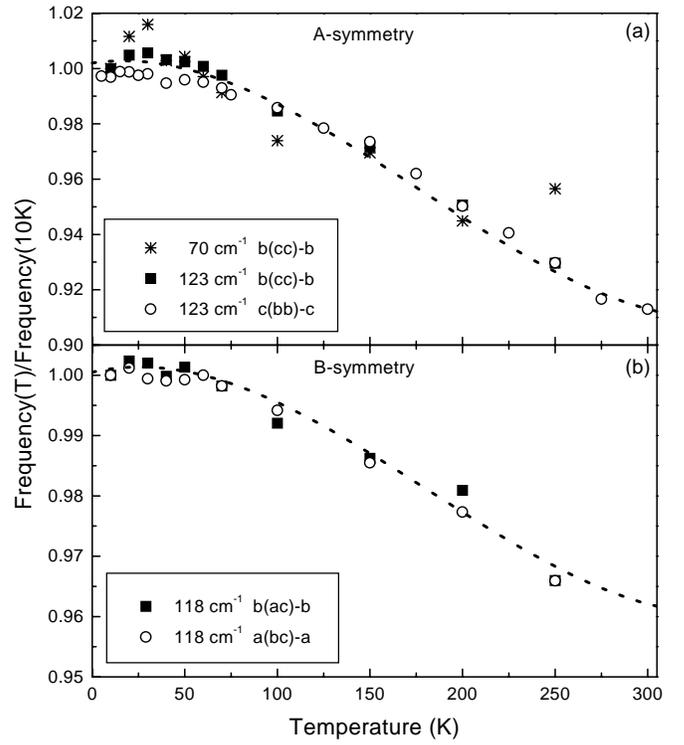

FIG. 3. Normalized frequency shift of the phonons. (a) 70 cm$^{-1}$ and 123 cm$^{-1}$ in $(bb)$ and $(cc)$ polarization, (b) 118 cm$^{-1}$ in $(ac)$ and $(bc)$ polarization. The dotted lines are guides to the eye.

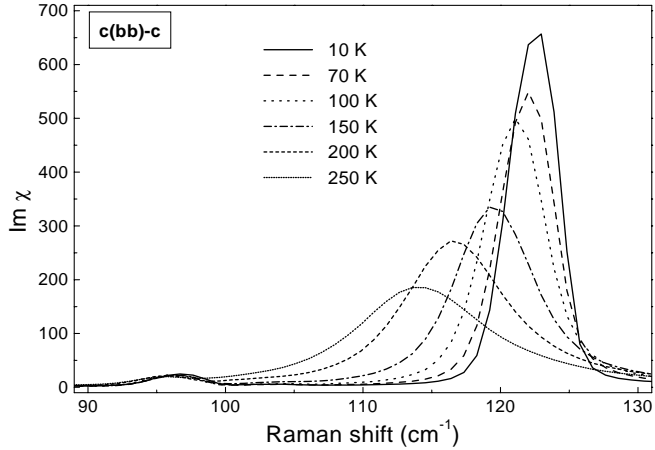

FIG. 2. Phonon anomalies in the Raman spectra of $(VO)_2P_2O_7$. As an example we present spectra of the mode at 123 cm$^{-1}$ in intrachain $(bb)$ polarization for different temperatures. In contrast to this scattering geometry the integrated intensity in the interchain $(aa)$ polarization shown in Fig. 4 is not preserved as function of temperature.

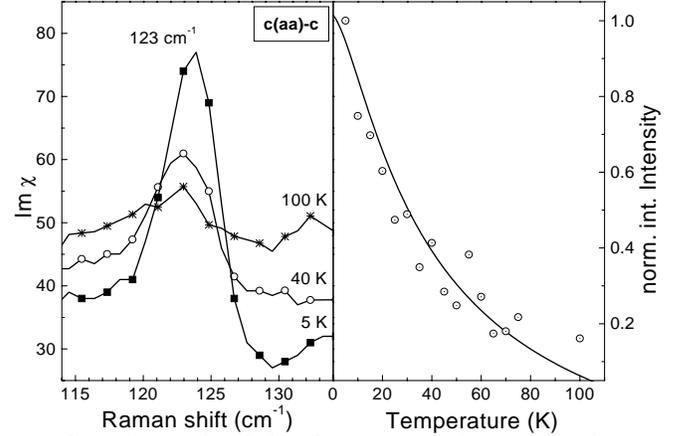

FIG. 4. Anomaly of the phonon scattering matrix element in interchain $(aa)$ polarization of the phonon at 123 cm$^{-1}$ as function of temperature (left panel). This phonon possesses a symmetric Lorentzian line shape. The intensity of the mode at 123 cm$^{-1}$ is shown in the right panel. The straight line corresponds to a fit with $I \propto 1 - \exp(-\Delta/T)$, $\Delta = \Delta_{01} = 36$ K.

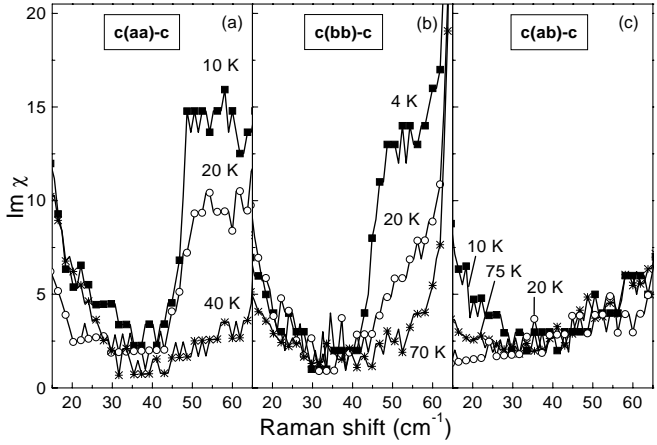

FIG. 5. Low frequency Raman scattering on $(VO)_2P_2O_7$ in the ab plane. The shoulder at 47 cm$^{-1}$ (onset at 45 cm$^{-1}$) is observed at low temperatures ($T < 75K \approx 2\Delta_{01}$) only with polarization both parallel to the alternating chain ($bb$) or both perpendicular ($aa$). In crossed polarization no magnetic light scattering at low energies is observable.

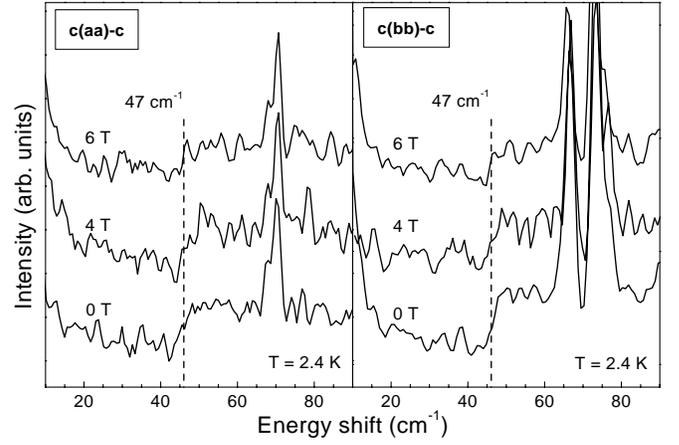

FIG. 7. Low frequency scattering of $(VO)_2P_2O_7$ in ($aa$) and ($bb$) polarization at $T = 2.4$ K for magnetic fields between 0 T and 6 T parallel to the a and b axis respectively. There is no splitting nor a shift of the shoulder at 47 cm$^{-1}$ observable.

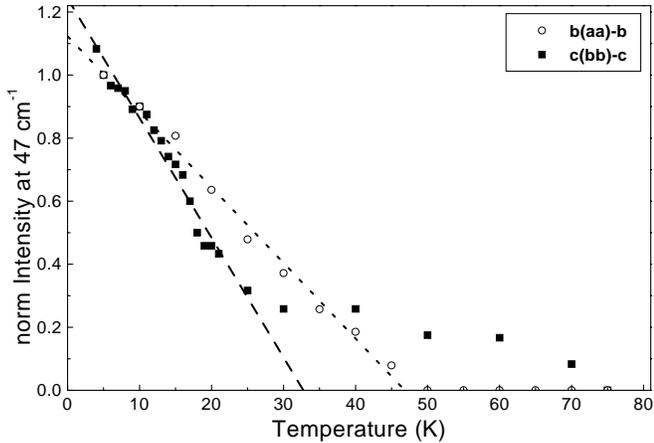

FIG. 6. Intensity of the gap-related scattering in $(VO)_2P_2O_7$ as function of temperature. For ($aa$) and ($bb$) polarization a linear increase is observed. Two temperature ranges can be clearly distinguished in intrachain ($bb$) polarization. The lines are guides to the eye.

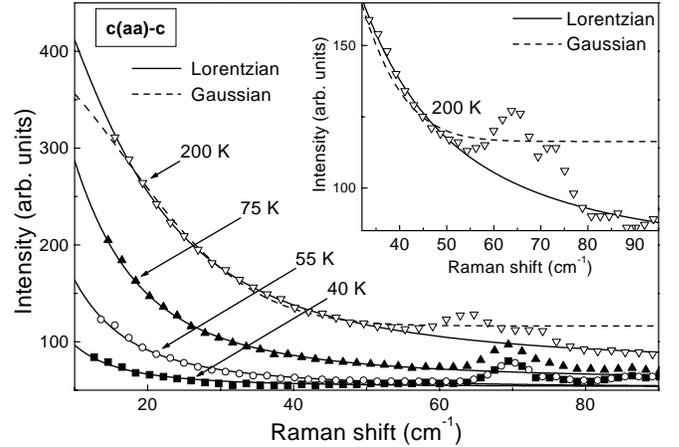

FIG. 8. Quasielastic scattering for ($aa$) polarization for temperatures between 40 K and 200 K. The solid and dashed curves denotes Lorentzian and Gaussian curves. The inset shows the Raman spectra at 200 K in comparison to a Gaussian and Lorentzian curves. The deviation between the Gaussian curve and the experimental data increases for energies greater than 45 cm$^{-1}$.

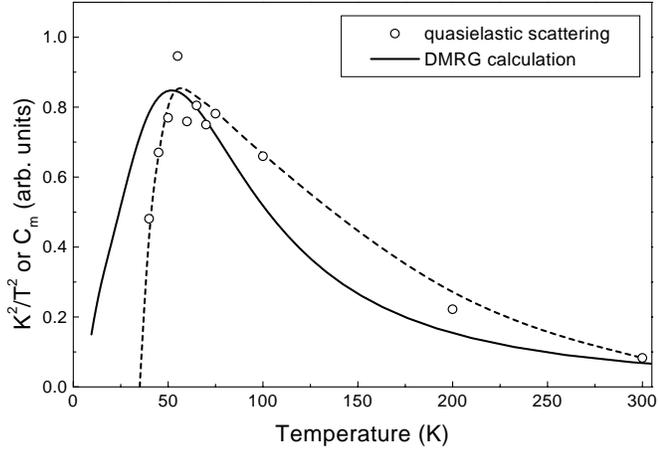

FIG. 9. The magnetic specific heat derived from the quasielastic scattering after Eq. 2 and Eq. 6. The dashed line is a guide to the eye. Note that the maximum in the magnetic specific heat is at $T_{max} = 55$ K. The solid curve corresponds to a DMRG calculation[35] of an alternating Heisenberg chain with a coupling constant $J = 110$ K and a dimerization $\delta = 0.09$.